\documentclass{kapproc} 

\kluwerbib

\begin{document}





\articletitle{The Dynamical Interaction of AGN with their Galaxian Environments}



\author{Michael A. Dopita}
\affil{Research School of Astronomy \& Astrophysics \\
The Australian National University}

\email{Michael.Dopita@anu.edu.au}







\begin{abstract}

Jet-driven shocks are responsible for an important fraction of the emission of the 
narrow-line regions (NLRs) in many classes of AGN. However, this cannot explain 
all observations. It is clear that the remaining sources are photoionised by
the active nucleus. The 2-d hydrodynamic models from the RSAA group support an
evolutionary scenario whereby the shock-excited NLRs are initially jet-driven 
but later, ionizing photons from the central engine replace shocks as the main
excitation mechanism and shock induced star formation may also become important.  
In their photoionized phase, dusty and radiation-pressure dominated evolution 
produces a self-regulated NLR spectrum. This model aso explains the coronal 
emission lines and fast (3000 km~s$^{-1}$) outflows seen in some Seyferts.

\end{abstract}





\section {Shock-Driven Radio Galaxies}

There is increasing evidence that the narrow-line regions (NLRs) associated
with many classes of active galactic nuclei (AGN) have a complex dynamical 
and excitation evolution. Evidence for a cocoon of strong, auto-ionising
and radiative shocks (see the theory  by \cite{Dopita95}, \cite{Dopita96}
and \cite{Bicknell97}) is particularly compelling for many luminous classes 
of radio galaxies. 
These include the steep-spectrum  radio sources (CSS), \cite{Fanti90}, unbeamed 
Gigahertz-peaked sources (GPS) (see the recent review  by \cite{O'Dea98}),
compact symmetric objects (CSO) (\cite{Wilkinson94}, and  references 
therein) or compact double sources (CD) \cite{Phillips82}. Together, 
these represent an appreciable fraction (10-30\%) of luminous radio 
sources. Not only are such sources very luminous at  radio frequencies, 
but they also are very luminous in optical emission lines, and spectra by 
Gelderman and Whittle (1994,1996) reveal intense ``narrow line'' emission 
with line ratios similar to those of Seyfert 2 galaxies.  For these 
objects, the line emission scales with radio power, and the continuity of 
properties across these different classes of sources argues strongly 
that the kinetic energy supplied by the radio-emitting jets may 
provide a substantial fraction of the power radiated at other
wavelengths by shocked gas associated with the NLRs of these galaxies.

The high redshift radio galaxies present us with a statistically significant
sample in which to study UV line ratio behaviour and so investigate the
fraction of the NLR emission produced by shocks, and the fraction by
photoionization. The study by Best et al. (2000a,b) revealed an extraordinary
result for powerful 3C radio galaxies with $z\sim 1.$ They find that both
the UV line profiles and the UV\ line ratio diagnostics imply that, when the
scale of the radio lobes is such that they are still able to interact with
the gas in the vicinity of the galaxy, they are predominantly shock-excited, 
but when the lobe has burst out into intergalactic space, the ionised gas 
left behind is predominantly photoionized. The ratio of fluxes in the
different classes of source suggests that the energy flux in the UV radiation 
field is about 1/3 of the energy  flux in the jets. Thus, both shocks and 
photoionization are important in the overall evolution of radio galaxies. 
This result, if confirmed for radio sources in general, would prove that that 
the properties of the radio jet are intimately connected with the properties 
of the central engine.

Very distant radio galaxies have been recently studied by De Breuck (2000).
He finds that diagnostic diagrams involving C~IV, He~II and C~III] fit to the
pure photoionization models, but that the observed C~II]/C~III] requires there
to be a high-velocity shock present. He argues that composite models would
be required to give a self-consistent description of all the line ratios, and
that these may require a mix of different physical conditions as well.

A fine example of such a source is provided by the $z\sim 3.8$ radio galaxy 
4C 41.17  has recently been studied in detail by \cite{Bicknell00}. This object 
consists of a powerful ``double-double'' radio source embedded in a 
$190 \times 130$ kpc Ly$\alpha$ halo (\cite {Reuland02}) and shows strong evidence for 
jet-induced star formation at $3000~\rm M_\odot~yr^{-1}$ associated with the 
inner radio jet.  \cite{Bicknell00} constructed a model involving the 
interaction of a high-powered jet with an energy flux of 
$\sim 10^{46}$ergs~s$^{-1}$ interacting with a dense cloud. Likewise, the 
outer jet has imposed a strong dynamical signature of outflow with velocities
in excess of 500 km~s$^{-1}$ in the line-emitting gaseous halo.

On the basis of such observations, we can propose a simple evolutionary
scenario for such radio sources. First, the accretion onto the central engine 
drives a radio jet which is first visible as a GPS source, but later 
evolves into a powerful 3C-like double lobe radio source. During the time 
that the scale of the radio lobes is less than 10-30~kpc, the interactions with 
the surrounding galactic medium are strong, and the NLR is predominantly 
shock-excited. The radio jets bore out ``ionization cones'' which are 
responsible for the ``alignment effect'' seen in the NLR of such sources. 
At late phases, the ionized gas is either photoionized by the central
source, or by shock-induced star formation that must inevitably
take place along the boundaries of the old shocked cocoon.

\section{Photoionisation-Dominated AGN}

Dynamical signatures of strong shocks are apparent in only $5-10\%$ of Seyfert
(eg.  \cite{Whittle96}). Many of these are the more radio-luminous galaxies
includinding Mrk78 (\cite{Pedlar89}), NGC2992  (\cite{Allen99}) or Mrk 1066 
(\cite{Bower95}). The power requirements are modest, typically 10$^{41}$ and 
10$^{44}$ ergs~s$^{-1}$  {\it c.f} those of luminous radio sources 
(10$^{45}$ - 10$^{46}$erg~s$^{-1}$). The remainder of Seyfert galaxies appear 
to be photoionised (\cite{Evans99}).

In the gas-rich and cloudy circumnuclear environments, light 
and low-power radio jets are readily disrupted and suffer entrainment from
the surrounding material, and molecular clouds are crushed in the high-pressure 
environment. This is clearly demonstrated by 2-d hydrodynamic simulations
by \cite{Bicknell02}, and in preparation. Here, shock 
velocities are generally lower and although shocks  may be still 
important in shaping the circum-nuclear medium, photoionisation is 
more important for its excitation.

A curious feature of the spectra of the NLR of photoionised 
Seyfert~2 galaxies is that nearly all are located in a region showing less 
than 0.8~dex variation in [OIII] $\lambda 5007$\AA$ /$H$\beta $, 
[NII] $\lambda 6583$\AA$ /$H$\alpha $ or [OI] $\lambda 6300$\AA$ /$
H$\alpha $ ratios, according to extensive observations by \cite{Veron00}. Within
individual galaxies, spatial variations in these line ratios are even tighter 
(\cite{Allen99}). The observations constrain the dimensionless ionization
 parameter, $U$, in the range $-3 < logU <-2$. This would require that 
the density of the photoionized  clouds falls off roughly as the inverse 
square of their radial distance.

Theoretical insight into this problem has recently been forthcoming 
(\cite{Dopita02}). Because clouds lying in the path of the jet and its 
surrounding high-pressure cocoon are crushed at relatively low velocity, then
any dust mixed with the cloud gas is likely to survive. If the central source 
produces UV photons with high local ionisation parameter, the dusty ionised
gas is compressed, raising the pressure close to the ionisation front
to match the radiation  pressure in the EUV radiation field. The regulates 
the apparent ionisation parameter, and ensures that the density of 
photoionized clouds varies as $R^{-2}$.  Each photoablating cloud is 
surrounded by a coronal medium in which the local ionization parameter 
reflects the ``true'' ionisation parameter delivered by the central 
source, and each has a dusty photo-accelerated radial tail  
(c.f. \cite{Cecil02}). In this model, the terminal velocity of outflow 
should correlate with the strength of the coronal line emission, 
explainig the observations of \cite{Zamanov02} and \cite{Rodriguez02}.





\begin{acknowledgments}

M. Dopita acknowledges the support of the ANU and  the Australian 
Research Council through his ARC Australian Federation Fellowship, 
and under the ARC Discovery project DP0208445.

\end{acknowledgments}






%






\bibliographystyle{apalike}

\chapbblname{}

\chapbibliography{}


\end{document}